 \documentclass[prb,twocolumn,showpacs,amsmath,amssymb,superscriptaddress]{revtex4}
 \usepackage{graphicx}% Include figure files
\usepackage{dcolumn}% Align table columns on decimal point
\usepackage[sort&compress]{natbib}
\usepackage{subfigure}
\usepackage{ifpdf}
\usepackage{bm}
\ifpdf
\usepackage[pdftex,
        colorlinks=true,
        pdftitle={},
        pdfauthor={S. Salem-Sugui Jr.},
        pdfsubject={fluctuation in iron-pnictides},
        pdfkeywords={IF, UFRJ, Rio de Janeiro Brazil},
        baseurl={http://www.if.ufrj.br},
        pdfpagemode=UseNone,
        bookmarksopen=true
        pdfoutput=1
        ]{hyperref}
\fi
\begin{document}
\title{Fish-tail and vortex dynamics in Ni doped iron-pnictide  $BaFe_{1.82}Ni_{0.18}As_2$ }
\author{S. Salem-Sugui Jr.}
\affiliation{Instituto de Fisica, Universidade Federal do Rio de Janeiro,
21941-972 Rio de Janeiro, RJ, Brazil}
\author{L. Ghivelder} 
\affiliation{Instituto de Fisica, Universidade Federal do Rio de Janeiro,
21941-972 Rio de Janeiro, RJ, Brazil}
\author{A. D. Alvarenga}
\affiliation{Instituto Nacional de Metrologia Normaliza\c{c}\~ao e
Qualidade Industrial, 25250-020 Duque de Caxias, RJ, Brazil.}
\author{L.F. Cohen}
\affiliation{The Blackett Laboratory, Physics Department, Imperial College London, London SW7 2AZ, United Kingdom}
\author{ Huiqian Luo}
\affiliation{National Lab for Superconductivity, Institute of Physics and National Lab for Condensed Matter Physics, P. O. Box 603 Beijing, 100190, P. R. China}
\author{Xingye Lu}
\affiliation{National Lab for Superconductivity, Institute of Physics and National Lab for Condensed Matter Physics, P. O. Box 603 Beijing, 100190, P. R. China}
\date{\today}
\begin{abstract}
We study the vortex-dynamics of a   $BaFe_{1.82}Ni_{0.18}As_2$ crystal with Tc = 8 K, by measuring flux-creep over the second magnetisation (or fish-tail) peak  for both H$\parallel$c-axis and H$\parallel$ab planes. Magnetic relaxation data show an anomalously long initial stage of relaxation, lasting approximately10 minutes for H$\parallel$c-axis and 2-3 min for H$\parallel$ab, resembling a transient effect with a lower relaxation rate, which is followed by the usual log(time) relaxation. Interestingly, study of the relaxation rate R vs H for both stages of relaxation and for both field directions, are featureless over the field range associated with the fish-tail.  The same trend was confirmed by plotting R vs T obtained from flux-creep data measured as a function of temperature for a fixed field (H$\parallel$c-axis). A plot of the activation energy U(M,T) calculated from the time relaxation of the magnetisation at a fixed field  also shows a smooth behavior further supporting the view that  the fish-tail peak is not associated with a crossover in vortex pinning regime within the collective pinning scenario.
\end{abstract} 
\maketitle 
The observation and study of the second magnetisation or fish-tail peak in pnictides \cite{japan} has attracted the attention of an increasing scientific audience. The relatively large flux-creep observed in this material system allows detailed studies of  the vortex-dynamics and comparison to a rich variety of models \cite{yeshurunRMP}. Untill now the fish-tail effect has been studied in $SmFeAsO_{0.9}F_{0.1}$ \cite{wen3}, $NdFeAsO_{0.85}$ \cite{moore}, $BaFe_{2-x}K_xAs_2$ \cite{wen1,us}, $BaFe_{2-x}Co_xAs_2$  \cite{wen2,nakajima,apl,proz,physC,phaset}, and more recently in $LiFeAs$ \cite{pra} and $PrFeAsO_{0.60}F{0.12}$ \cite{bhoi}. Different mechanisms have been claimed to account for the fish-tail effect in these compounds (see for instance a table presented in Ref.\onlinecite{us}) evidencing the need to build up a more comprehensive picture through additional studies, particularly on systems not yet investigated.

In this work, we present a vortex-dynamics study performed on the recent synthesized Ni doped iron pnictide superconductor system  $BaFe_{1.82}Ni_{0.18}As_2$ \cite{Luo}. We study an overdoped single crystal  with  x=0.18,  $T_c$= 8 K ($\delta$$T_c$$\approx$1 K), m=78 mg, and average dimensions 1.1x0.4x0.025 cm, for magnetic fields applied along the ab-planes direction.  In this geometry, a possible misalignment of the sample with respect to the direction of the applied magnetic field is estimated to be smaller than 2 degrees,  which assures that any contribution of the c-axis component to the magnetization measured with H$\parallel$ab-planes is neglegible. A small piece with m=23.4 mg and average dimensions 04x0.4x0.02cm was broken from the large sample for the measurements with H$\parallel$c-axis.  ÊDetails on sample preparations and physical properties of the large high-quality single crystals  can be found in Ref.\onlinecite{Luo}. Isothermal M(H) curves for the crystals for H$\parallel$ab-planes and  H$\parallel$c-axis exhibit the fish-tail peak,  the maximum occurring at an applied field Hp. 
The vortex dynamic investigation was performed by measuring isofield and isothermic magnetic relaxation, M vs time curves, over a period of 1 hour for magnetic field values lying below and above Hp on both branches of the M(H) curves. All data was obtained after cooling the system from above $T_c$ to the desired temperature in zero applied magnetic field (but in the presence of the earth magnetic field), which is called the zfc procedure. Magnetization data was obtained using commercial magnetometers based on a superconducting quantum interference device (SQUID) (MPMS-5T Quantum Design, and Criogenics-6T). The charge and discharge rate of the magnet were set equal to 100 Oe/s during the experiment.  
\begin{figure}[t]
% Requires \usepackage{graphicx}
\includegraphics[width=\linewidth]{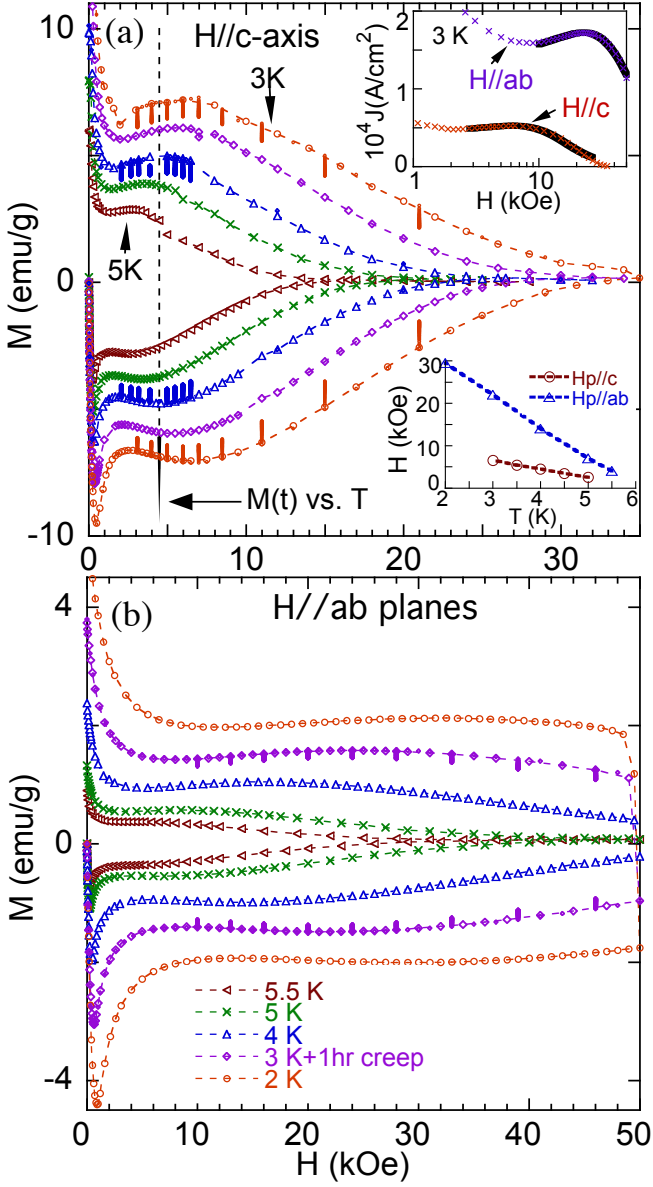}
\caption{Isothermic M(H) curves. a) H$\parallel$c-axis, at 3, 3.5,4, 4.5 and 5 K. The upper inset shows $J_c$(T=3K)vs H in logarithmic scale for both field directions. Solid lines in the region of the peak represents a fit  to an expression of Ref.\onlinecite{baruch}. The lower inset shows Hp vs T extracted from M(H) curves for both field directions; b) H$\parallel$ab-planes.}
 \label{fig1}
\end{figure}

Figure 1 shows isothermal M(H) curves exhibiting the fish-tail effect, as obtained for both geometries. M(H) curves at 3K and 4K for H$\parallel$c and at 3K for H$\parallel$ab are plotted with correspondent magnetic relaxation data obtained on both branches. We refer to this as the isothermal method. The dotted line located at H=4.6 kOe in Fig. 1a show that once this field value is fixed it is possible to go from below Hp to above it by increasing the temperature from 3 K. This second method, which we refer to as the isofield method, provides an independent check of the relaxation data taken using the isothermal route and allows the creation of an activation energy curve with temperature U(T) as in Ref.\onlinecite{maley,maley2}. As depicted in Figure 1, all M(H) curves are quite symmetric relative to the x-axis evidencing that vortex dynamics is mostly due to bulk pinning. Also, since the equilibrium magnetization, $M_{eq}$=$(M^++M^-)/2$$\approx$0, where $M^+$ and $M^-$ refer to the magnetization in the upper and lower branches of the hysteresis curve respectively, we use M instead $M-Meq$ in all curves. The upper inset of Fig.1 shows a plot of the critical current $J_c$ vs H at 3K for both geometries as estimated from the correspondent hysteresis curves shown in Fig.1 by using the well know expression $J_c=20\Delta M(emu/cm^3)/[a(1-a/3b)]$ where b$>$a   \cite{bean}.  The lower inset of Fig.1 shows a plot of the fish-tail peak field, Hp, as a function of temperature extracted from the curves of Fig.1 for both field directions, where dotted lines are only a guide for the eyes. We observe from the inset figures that $J_c$(Hp$\parallel$ab)/$J_c$(Hp$\parallel$c)$\approx$3.3 and Hp(H$\parallel$ab)/Hp(H$\parallel$c)$\approx$3.6. Although neither of these parameters is a fundamental measure of the superconducting anisotropy (for that we would have to extract the upper and lower critical fields), they clearly are consistent with each other, and the fact that the anisotropy expected for this system is of the order of 3 \cite{Ni}. 

Figure 2 show selected magnetic relaxation data obtained on both branches of the M(H) curve at 3 K for H$\parallel$c-axis. These curves exemplify the general behavior observed on all relaxation curves obtained in this work, namely, all curves show first a slow relaxation rate, and then an increase in the relaxation rate after a time $\tau_0$$\approx$10 minutes after relaxation starts for H$\parallel$c and,  $\tau_0$$\approx$2-3 minutes for H$\parallel$ab. This anomalous behavior showing two distinct linear behavior of M with log(t) or two distinct time windows, can not be explained in terms of a pinning crossover, since in that case the crossover always occur from a higher rate of the initial stage of relaxation (corresponding to flux jumping over low activation energy pinning sites) to a lower rate (corresponding to flux jumping over higher activation energy pinning sites), as occurring in the well known surface to bulk pinning crossover \cite{burlachkov}. This scenario is the opposite of what is observed here. There are two other possible explanations for the non-linearity of M(t) vs logt curves. The first is due to an eddy current induced on the sample by the ramp rate dH/dt while charging the magnet  \cite{gurevich}. After the magnet is charged and dH/dt=0 the induced eddy current decays due to flux-creep producing a transient region which possesses a lower magnetic relaxation rate when compared to that due to the bulk pinning.  The duration time of this transient relaxation is inversely proportional to the charge magnet rate dH/dt  and it is expected to last less than a minute for a rate dH/dt$\approx$100 Oe/s \cite{gurevich} as used here. This effect referred to as an initial settle time, is commonly observed in M(t) curves, and is inversely proportional do the ramp rate of the magnetic field \cite{gurevich} and found to be of the order of 1 minute usually.
It should be mentioned that the ratio between the times $\tau_0$(H$\parallel$c)/$\tau_0$(H$\parallel$ab)$\approx$4 for our data qualitatively agrees with the hypothesis that the anomalous initial decay of M(t) is related to the transient relaxation associated to dH/dt, once $\tau_0$$\approx$$J_1a/(dB/dt)$\cite{gurevich} where $J_1$ represents the eddy current (J(H$\parallel$ab)$\approx$3J(H$\parallel$c see Fig. 1), $a$ is the sample thickness perpendicular to H ($a$=0.025cm for H$\parallel$ab and 0.4cm for H$\parallel$c), and dB/dt is the rate of the magnetic induction field B. The other possible explanation, is based on the existence of twin-boundaries \cite{maple} which may allow a continuous increase of vortices pinned between twin-boundaries (appearing in the first time window). As a result, after the density of vortices pinned by twin-boundaries increases above a certain value, the pressure overcomes the barrier producing an increase in the rate of relaxation at the second time window \cite{maple}. This explanation would apply as well to our experiment, since the efficiency of twin-boundaries, as pinning sites, changes and become less effective as the sample rotates from  H$\parallel$c to H$\parallel$ab (this because many TB aligned parallel to the fied when H$\parallel$c can become perpendicular to the field after the 90 degree rotation). However, after careful observation of the sample surfaces in a polarized light microscope, there are no visible twin-boundaries, and up to this point, twinning has only been observed in underdoped not overdoped crystals.  Consequently we assign the initial relaxation to a transient effect.
 \begin{figure}[t]
% Requires \usepackage{graphicx}
\includegraphics[width=\linewidth]{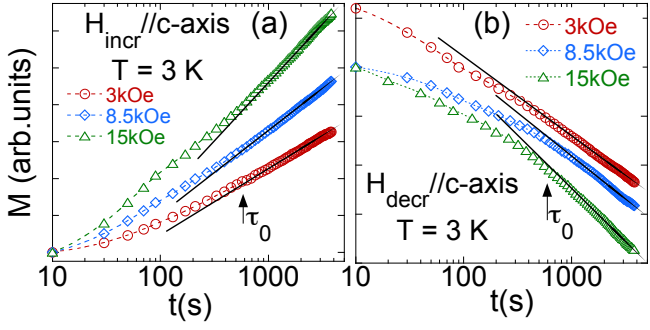}
\caption{Selected M vs time curves for H$\parallel$c:  a) lower branch of M(H). b) upper branch of M(H). Solid lines are only a guide to the yes evidencing the increasing relaxation rate for t$>$$\tau_0$.}
 \label{fig2}
\end{figure}

\begin{figure}[t]
% Requires \usepackage{graphicx}
\includegraphics[width=\linewidth]{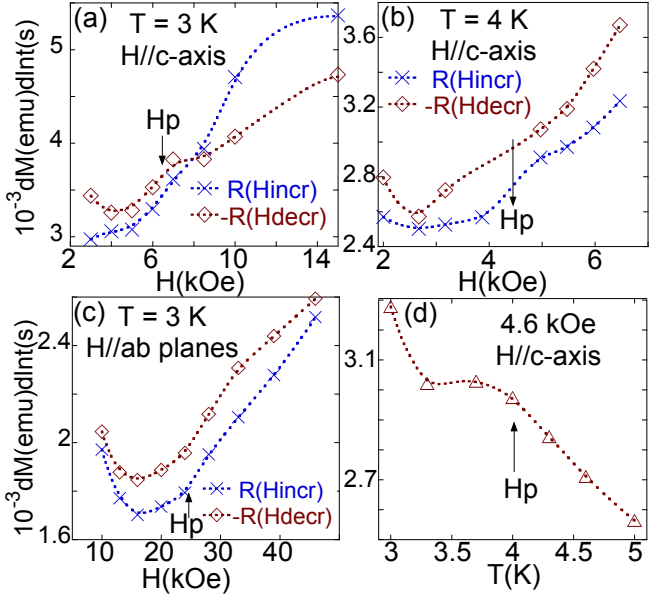}
\caption{Plots of the relaxation rates; R vs H for both branches of M(H) curves for: (a) and (b) H$\parallel$c-axis; (c) H$\parallel$ab; (d) R vs T for H=4.6 kOe, H$\parallel$c.}
 \label{fig3}
\end{figure}
We analyze flux-creep data by obtaining the relaxation rate $R=dM/dlnt$ of the second stage of relaxation (corresponding to the region t$>$$\tau_0$) for each M(t) curve \cite{beasley,donglu,lesley1,lesley2,tb}. Figure 3 show the results from relaxation obtained in the lower branch ($H_{incr}$) and in the upper branch ($H_{decr}$) of M(H) curves for both geometries.  We mention that we also obtained values of R(t$<$$\tau$) for the first stage of relaxation, and observe that plots of R(t$<$$\tau$)vs H approximately follow the same trend as the plots in Fig.3. This fact may evidence that the same major bulk pinning mechanism is present in both stages of relaxation, supporting our hypothesis that the anomalous first stage of relaxation is related to a transient effect rather than to a twin-boundary assisted effect. The main information one can extract from the many plots of Fig. 3 is the absence of any feature, minimum or maximum, located near the correspondent fish-tail peak position represented by a vertical arrow in each figure. This fact strongly suggests that the fish-tail peak observed in M(H) curves for both field directions is not due to a pinning crossover. 

\begin{figure}[t]
% Requires \usepackage{graphicx}
\includegraphics[width=\linewidth]{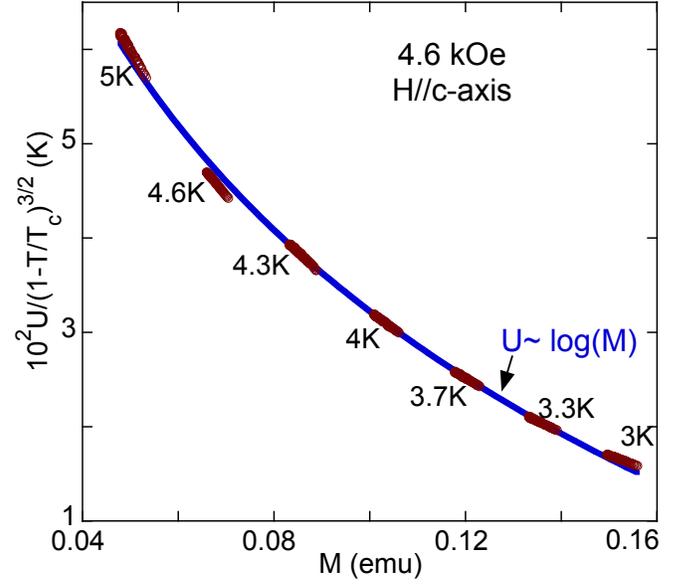}
\caption{The scaled activation energy U(M,T) for H=4.6 kOe are plotted against M. The solid line is a logM fit.}
 \label{fig4}
\end{figure}

We also obtain flux-creep data from the isofield method, with a path represented in Fig. 1a by a vertical dotted line for H=4.6 kOe. M(t) data were obtained by reaching the desired temperature in a zfc procedure, followed by an increase of the field up to 4.6 kOe and measuring M(t) for 60 min. Results of R=dM/dlnt obtained this way for all M(t) data is presented in Fig. 3d. As in other plots of R vs H shown in Fig. 3, the plot of Fig. 3d does not show any marked feature as Hp (located at T=4K) is crossed. 
The isofield activation energy curve U(M,T), or $U(M,T)/g(T/T_c)$ where $g(T/T_c)$ is some scaling function, is believed to be a smooth function of M, where its behavior with M can provide information on the pinning mechanism \cite{beasley,maley,maley2}.  Here we performed an analysis of the activation energy as shown previously in Ref.\onlinecite{maley,maley2} using the expression $$U=-Tln(dM(t)/dt)+CT$$ where C is a constant which depends on the hoping distance of the vortex, the attempt frequency and the sample size. The constant C is adjusted in a manner such that all $U(M,T)/g(T/T_c)$ curves  plotted against M fall on a smooth curve. If all data does not fall on one smooth curve with a fixed value of C, it is an indication that the data does not fall into one pinning regime. Figure 4 show the results of the analysis. The appropriate scaling function for our case is $g(T/T_c)$=$(1-T/T_c)^{3/2}$ also used in Ref.\onlinecite{maley2} and the constant C=14; similar values of the constant C have been found for high-$T_c$ cuprates \cite{maley,maley2} and pnictides \cite{us}. The plot of U(T) show an almost perfect logarithmic with M, indeed suggesting that a single pinning regime is operating in the temperature range studied at the fixed field of 4.6 kOe. 

As a pinning regime crossover does not appear to be an appropriate explanation for the fishtail effect it is interesting to explore whether the data can be understood within the model proposed for a vortex lattice phase transition \cite{baruch}, using the expression $J_c(B)$$=$$A/[(B-Bp)^2+(\Delta B)^2]^{5/4}$ where A is a fitting parameter and $\Delta B$ is the peak width. The M vs H curves have been fit to this expression, converting the magnetisation data to critical current as described previously and  using  B$\approx$H. The good-quality fittings conducted on $J_c$(T=3K) for both field directions are shown as solid lines in the curves of the upper inset of Fig. 1a. By considering $J_c$=$\Delta$M(emu/$cm^3$) the values of the fitting parameters are: A=1.1x$10^4$ $G^{5/2}$ and $\Delta B$=14.2 kOe for H$\parallel$c, and A=4.4x$10^4$ $G^{5/2}$ and  $\Delta B$=46 kOe for H$\parallel$ab.  Note that the fittings produced larger values for the peak width compared to values obtained in Ref.\onlinecite{phaset}. It is intersting that the data fits well to this expression, although we recognize that this in itself is not conclusive evidence of a phase transition in the vortex lattice. Moreover we would expect a change in vortex dynamics as a signature that a phase transition has occurred, and we see no evidence for this. 

In conclusion, our experiments show that the fish-tail peak appearing on M(H) curves for  H$\parallel$c and H$\parallel$ab in  $BaFe_{1.82}Ni_{0.18}As_2$ does not appear to be associated with a softening in vortex pinning prior to melting, nor a change of pinning regime within a collective pinning model scenario. We observe an anomalously long relaxation period at the initial stage of relaxation lasting $\approx$10 minutes for H$\parallel$c which is likely to be a transient effect due to the field rate dH/dt while charging (discharging) the magnet. This transient effect appears to be a feature intrinsic to the BaFeNiAs system we have studied, as we have also found it  present in preliminary data obtained in a second crystal with x=0.1 ($T_c$$\approx$20K) for H$\parallel$ab-planes.

SSS, LG and ADA thanks support from the Brazilian agencies CNPq and FAPERJ. LFC thank the UK Funding Council the EPSRC grant EP/H040048.

\end{document}